\documentclass[twocolumn,english,NOlinenumbers,aps,twocolumn,groupedaddress,superscriptaddress]{revtex4-1}
\usepackage[T1]{fontenc}
\usepackage[utf8]{inputenc}
\usepackage{color}
\usepackage{babel}
\usepackage{graphicx}
\usepackage{esint}
\usepackage[unicode=true,pdfusetitle,
 bookmarks=true,bookmarksnumbered=false,bookmarksopen=false,
 breaklinks=false,pdfborder={0 0 1},backref=false,colorlinks=true]
 {hyperref}
\hypersetup{
 citecolor=blue}

\makeatletter
\usepackage{babel}

\makeatother

\begin{document}

\title{Arrest of Domain Coarsening via Anti-periodic Regimes in Delay Systems }

\author{J. Javaloyes}

\affiliation{Departament de Fisica, Universitat de les Illes Baleares, C/ Valldemossa
km 7.5, 07122 Mallorca, Spain.}

\author{T. Ackemann}

\affiliation{SUPA and Department of Physics, University of Strathclyde, John Anderson
Building, 107 Rottenrow, Glasgow G4 0NG, UK.}

\author{A. Hurtado}

\affiliation{Institute of Photonics, SUPA and Department of Physics, University
of Strathclyde, Technology and Innovation Centre, 99 George Street,
Glasgow G1 1RD, UK}
\begin{abstract}
Motionless domains walls representing heteroclinic temporal or spatial
orbits typically exist only for very specific parameters. This report
introduces a novel mechanism for stabilizing temporal domain walls
away from the Maxwell point opening up new possibilities to encode
information in dynamical systems. It is based on anti-periodic regimes
in a delayed system close to a bistable situation, leading to a cancellation
of the average drift velocity. The results are demonstrated in a normal
form model and experimentally in a laser with optical injection and
delayed feedback. 
\end{abstract}
\maketitle
Spatially extended nonlinear systems often admit multiple coexisting
stable states, and the dynamical properties of the fronts connecting
them are fundamental in the understanding of pattern formation and
localized structures (LS). These dissipative objects \cite{FT-PRL-90,AA-LNP-05,AA-LNP-08,UMS-NAT-96,AP-PLA-01,AAP-PRL-97,NAD-PSS-92,LMP-NAT-94,WKR-PRL-84,MFS-PRA-87,1172836}
occur in many natural and laboratory systems and are characterized
by a correlation range much shorter than the size of the system, making
them individually addressable. Localized structures due to a front
between a homogeneous and a cellular state are intrinsically stable
\cite{pomeau86,CRT-PRL-00,CRT-C-04}, as the pattern oscillations
stabilize the position of the front against small perturbations. In
contrast, fronts between homogeneous states are typically stable only
for very specific parameters and lead otherwise to a coarsening.

In this letter, we discuss a robust mechanism for the stabilization
of fronts between homogeneous states based on a generic property of
time delayed systems having no equivalent in spatially extended systems.
We show that structures occurring within a delay time can be stabilized
by a temporal inversion after each time delay, thus allowing the stabilization
of self-localized temporal domains in temporal slices modulo twice
the delay. The case of equispaced domain walls reduces to the so-called
square-waves \cite{MPN-AMPA-86,N-PD-03}. This has not only relevance
for the fundamentals of the dynamics of delayed system, but also meets
recent interest in temporal LS realized in photonics \cite{LCK-NAP-10,TG-OL-10,TBC-PRA-13,HBJ-NAP-14,MJB-PRL-14,MJB-JSTQE-14,GJT-NC-15}.
Due to their intrinsically fast dynamics, the possibility to use LS
as bits for information processing was addressed early in nonlinear
optics \cite{rosanov,TML-PRL-94,FS-PRL-96}. Interesting results were
achieved for spatial solitons in semiconductor microcavities \cite{BTB-NAT-02,GBG-PRL-08,TAF-PRL-08},
although it turned out that spatial disorder limits parallelism and
control \cite{PTB-APL-08,ARN-OL-12}, motivating further studies into
temporal LS.

As indicated, in the simple case of a unidimensional bistable system
with a single dynamical variable $\psi\left(x,t\right)$, the stable
coexistence between two homogeneous phases is merely achieved for
a single parameters' value, the so-called Maxwell point. Here a domain
wall separating the two states would be motionless. Yet, such a regime
possesses little experimental significance since any deviation of
the control parameter or any symmetry breaking effect implies that
one of the two bistable phases will eventually invade the other in
a way reminiscent of nucleation bubbles in first order phase transitions.

In recent years, building on the strong analogies between spatial
and delayed systems \cite{AGL-PRA-92,GP-PRL-96,K-CMMP-98}, a similar
symmetry breaking induced coarsening dynamics was shown to occur in
delayed bistable systems \cite{GMZ-EPL-12,GMZ-PRE-13}. The ability
to control the motion of these walls would have significant implications
as for instance to encode and process information with a fast nonlinear
delayed system. Motivated by this idea, the pinning of domain walls
was recently demonstrated via an external temporal modulation \cite{MGB-PRL-14}.
Departing from the analogies with spatial systems, we demonstrate
in contrast a stabilization mechanism based not upon a fast active
modulation, but upon a slow, self-induced dynamics. We envision the
use of a lesser known property of delayed systems: their ability to
generate anti-periodic output, i.e. temporal traces getting inverted
after each time delay $\tau$, thereby inducing an effective periodicity
$2\tau$. 

While this effect was studied theoretically in \cite{MPN-AMPA-86,N-PD-03,N-PRE-04}
and square-waves were demonstrated experimentally in several optical
\cite{GES-OL-06,MGJ-PRA-07} and opto-electronic systems \cite{WED-PRE-12},
we demonstrate in this letter how this generic property of anti-periodicity
can be harnessed to create robust motionless domain walls and prevent
the coarsening phenomenon. This allows to store information even far
from the Maxwell point and/or in the absence of bistability. Such
an idea has no equivalent in bona fide spatial systems since the anti-periodicity
would actually correspond to a space defined over a Möbius strip.
We evidence experimentally and theoretically stable domains in an
injected semiconductor laser with delayed feedback, that are\emph{
insensitive} to symmetry breaking and exist beyond the bistability.

We base our analysis on the normal form of the imperfect pitchfork
bifurcation modified by the inclusion of a linear delayed contribution
\begin{eqnarray}
\varepsilon\frac{d\Psi}{dt} & = & \mu\Psi+\beta\Psi^{2}-\Psi^{3}+\eta\Psi\left(t-1\right).\label{eq:PFDDE}
\end{eqnarray}
In Eq.~(\ref{eq:PFDDE}), the parameter $\mu$ controls the bistability,
$\beta$ represents the symmetry breaking and $\eta$ is the amplitude
of the time delayed feedback. The temporal scale is normalized by
the time delay $\tau$. We study the limit of long delays $\tau\rightarrow\infty$,
and define a smallness parameter $\varepsilon=1/\tau$ making apparent
the singular nature of Eq.~(\ref{eq:PFDDE}). Delay Differential
Equations (DDEs) possess in the long delay limit an eigenvalue spectrum
that can be divided in two parts, see for instance \cite{Y-DCDS-15}
and references therein. A quasi-continuous branch stems from the influence
of the delayed contribution while a discrete spectrum is generated
by the instantaneous linear terms. We are interested in the regimes
where a portion of the quasi-continuous branch may become unstable
giving rise to smooth dynamics. We set $\mu<0$ ensuring the stability
of the discrete spectrum. When $\varepsilon\rightarrow0$, the left
hand side of Eq.~(\ref{eq:PFDDE}) can be assumed, as a first approximation,
to vanish. Such an approach is enlightening as one finds a functional
mapping governing the evolution of the small deviations from the trivial
solution $\Psi=0$ as 
\begin{eqnarray}
\Psi\left(t\right) & = & -\kappa\Psi\left(t-1\right)\label{eq:Mapping}
\end{eqnarray}
with $\kappa=\eta\mu^{-1}$. Due to the infinite dimensionality of
DDEs, an initial condition must be given as a function defined over
an interval equal to the delay. If $|\kappa|<1$, any initial condition
slowly decays from one round-trip to the next and the steady state
$\Psi\left(t\right)=0$ is asymptotically stable. Yet, for $|\kappa|>1$,
the trivial solution may bifurcate via two widely different scenarios.
If $\kappa<-1$, the temporal profile evolves regularly from one round-trip
to the next while for $\kappa>1$, the profile gets inverted at each
round-trip signaling the onset of a period-2 (P2) regime.

The slowly evolving dynamics of the temporal profile can be better
understood and visualized via the spatio-temporal equivalence between
delayed and spatially extended systems \cite{AGL-PRA-92,K-CMMP-98,GP-PRL-96,N-PD-03,Y-DCDS-15}.
We define the deviation from the two bifurcation points $\mu=\kappa\eta$
with $\kappa=\pm1$, as $\mu=a+\varepsilon^{2}a_{1}$ and $\eta=a\kappa+b_{1}\varepsilon^{2}$
and, as detailed in \cite{Y-DCDS-15}, inserting a multiple time scales
expansion in Eq.~(\ref{eq:PFDDE}) for both the temporal derivative
and the delayed term, and defining $\Psi=\varepsilon\psi+\mathcal{O}\left(\varepsilon^{2}\right)$,
one obtains the following equivalent partial differential equation
(PDE) as a third order solvability condition
\begin{eqnarray}
\frac{\partial\psi}{dn} & = & p\psi+\frac{\beta}{a}\psi^{2}+\frac{\psi^{3}}{a}+\frac{1}{2a^{2}}\frac{\partial^{2}\psi}{\partial x^{2}},\label{eq:PFPDE}
\end{eqnarray}
with $p=-\left(a_{1}-\kappa b_{1}\right)/a$. Although strictly valid
in the vicinity of the bifurcation points, such normal forms are known
to have a wider domain of validity, see the discussion before Eq.~(8a)
in \cite{N-PD-03}. In Eq.~(\ref{eq:PFPDE}), the spatial coordinate
($x$) must be understood as a local time coordinate within the round-trip
while the slow time ($n$) represents the evolution of the temporal
profile from one round-trip to the next. Besides, we factored out
a drift velocity defined as $\upsilon=a^{-1}$ as in \cite{GP-PRL-96,GJT-NC-15},
representing the small deviation of the period with respect the time
delay. Equation~(\ref{eq:PFPDE}) must be complemented by a boundary
condition that reads 
\begin{eqnarray}
\psi\left(x+1,n\right) & = & -\kappa\psi\left(x,n\right)\label{eq:BC}
\end{eqnarray}
and defines, as in Eq.~(\ref{eq:Mapping}), whether or not the solution
gets inverted from one round-trip to the next. For $\beta=0$ and
$a<0$, the solutions of the heteroclinic orbits of Eq.~(\ref{eq:PFPDE})
read $K_{\pm}\left(x\right)=\pm\sqrt{\left|a\right|p}\tanh\left(\left|a\right|\sqrt{p}x\right)$
where the $\pm$ stands for the upward and downward domains walls.
In deriving Eq.~(\ref{eq:PFDDE}) we assumed that the symmetry breaking
term $\beta$ scales as $\varepsilon$ to enter as a perturbation
of the solvability condition. Yet, although small, $\beta$ has a
deep impact upon the dynamics. 

\begin{figure}
\begin{centering}
\includegraphics[bb=30bp 50bp 600bp 390bp,clip,width=1\columnwidth,height=6cm,keepaspectratio]{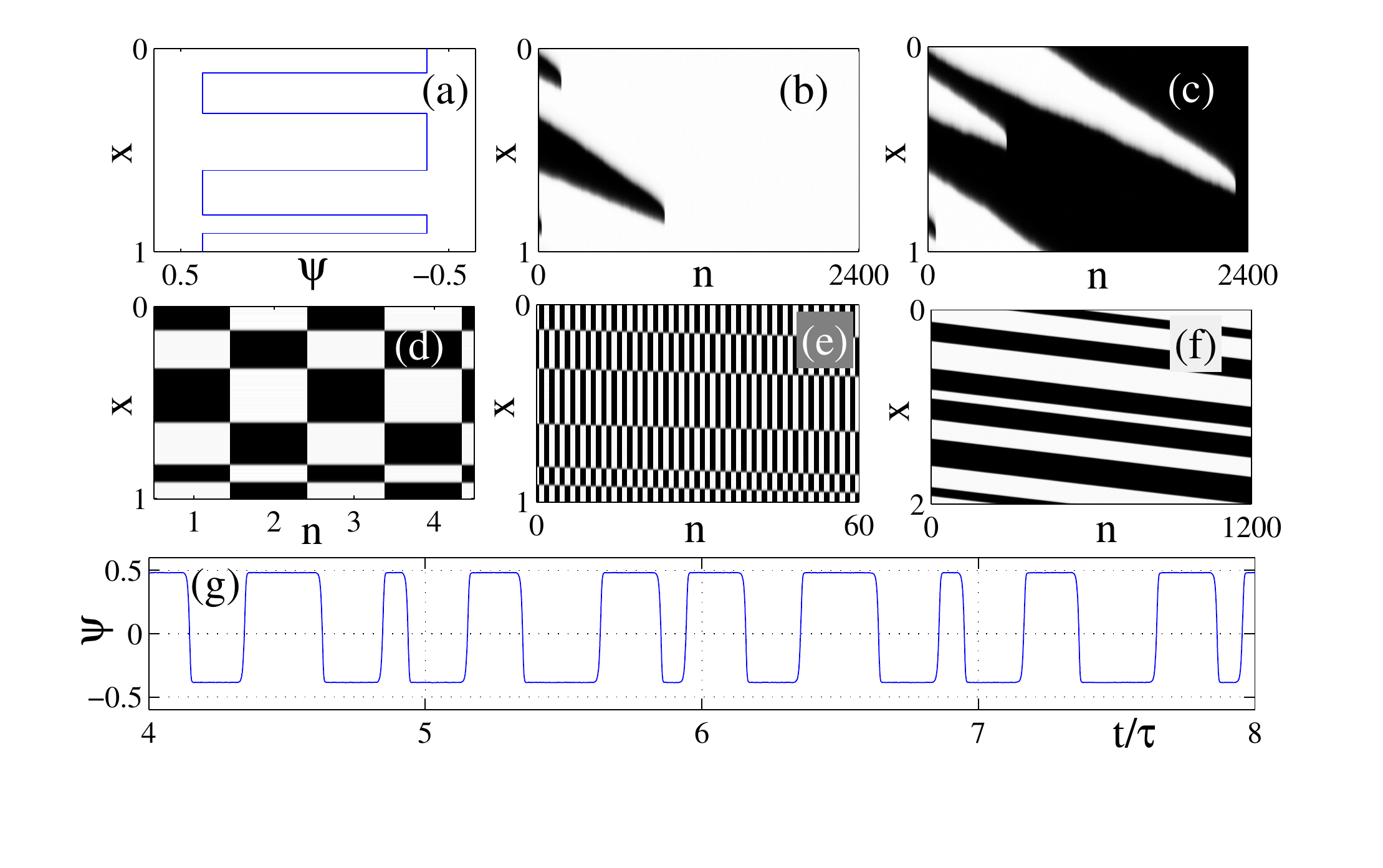}
\par\end{centering}

\caption{{\footnotesize{}(color online). Illustration of dynamics in space-time
diagrams. These space-time diagrams are folded over a time $T=\tau+|\eta|^{-1}$.
a) Temporal pattern imposed as an initial condition. The P1 regime
where $\kappa=-1$ leads to the coarsening dynamics of the temporal
time trace depicted in b) and c) towards the high and low states when
$\beta=-10^{-2}$ b) and $\beta=10^{-2}$ c), respectively. Other
parameters are $\tau=10^{3}$, $\eta=5\times10^{-2}$ and $\mu=2.5\times10^{-2}$,
or equivalently, $a=-5\times10^{-2}$, $b_{1}=0$ and $a_{1}=7.5\times10^{4}$.
The P2 regime where $\kappa=1$, induces the inversion of the initial
condition at each round-trip as exemplified by the checkerboard pattern
in d). It does not coarsen over longer time scales even for very large
values of the asymmetry $\beta=0.1$. The time trace is folded over
times $T$ in e) and also $2T$ in f) for clarity. Parameters are
$\eta=-0.2$ and $\mu=-2.5\times10^{-2}$, or equivalently $a=-0.2$,
$b_{1}=0$ and $a_{1}=17.25\times10^{4}$. The temporal profile is
represented in g). Notice the large values of $a_{1}$ signaling that
Eq.~(\ref{eq:PFPDE}) remains qualitatively valid far from the bifurcation
points. }}
\label{figtheo1}
\end{figure}

We depict in Fig.~\ref{figtheo1}a) an initial condition for Eq.~(\ref{eq:PFDDE})
composed of an arbitrary succession of domains with values corresponding
to the plateaus of the P1 or of the P2 regimes. In the P1 case ($\kappa=-1$)
as visible in Fig.~\ref{figtheo1}b,c), this multi-plateau pattern
relaxes in a finite time to the upper (resp. lower) value when $\beta<0$
(resp. $\beta>0$). Such a coarsening scenario is very general and
it is preserved quite far from the bifurcation condition $\mu=-\eta$,
and even when the discrete spectrum is also unstable, i.e. with $\mu>0$.
A multi-domain solution that verifies Eqs.~(\ref{eq:PFPDE},\ref{eq:BC})
can be written as 
\begin{eqnarray}
\psi\left(x,n\right) & = & K_{+}\left(x-x_{0}^{+}\right)+K_{-}\left(x-x_{1}^{-}\right)\\
 & + & \cdots+K_{-}\left(x-x_{n}^{-}\right)-\sqrt{\left|a\right|p}\nonumber 
\end{eqnarray}
with $\left\{ x_{n}^{\pm}\right\} $ the ordered collection of the
coordinates of the upward and downward kinks. For $|\beta|\ll1$,
a variational approach is justified \cite{C-IJBC-02}, and we insert
an ansatz as $\psi\left(x,s\right)=K_{\pm}\left[x-x_{\pm}\left(s\right)\right]$.
By multiplying Eq.~(\ref{eq:PFPDE}) by $\partial_{x}K_{\pm}$ and
integrating over the spatial coordinate, we find the effective equation
for the motion of an isolated wall as
\begin{eqnarray}
\frac{dx_{i}^{\pm}}{dn} & = & \mp\frac{\beta}{2a\sqrt{\left|a\right|}}
\end{eqnarray}
demonstrating that the symmetry breaking term $\beta$ induces a splitting
of the velocity of domain walls of opposed ``charge'', that will
eventually collide. Notice that the value of the drift can also be
found searching for heteroclinic solutions of Eq.~(\ref{eq:PFDDE})
as in \cite{GMZ-EPL-12}. Both methods neglect the short range, exponentially
decreasing interactions between nearby walls that is known to be attractive
\cite{C-IJBC-02,N-PD-03,N-PRE-04}. This can leads for $\beta<0$
to a steady state, that is however unstable and cannot prevent the
eventual collision and coarsening. Hence, one may conclude that it
is impossible to store information in a scalar delayed system as any
symmetry breaking nonlinearity leads to a coarsening of the information. 

We now demonstrate how a completely different regime can be obtained
exploiting the anti-periodic solutions of Eq.~(\ref{eq:PFDDE}) achieved
by simply setting $\eta<0$ to access the regime where $\kappa=1$.
We stress that here since $\mu<0$, bistability is lost and the only
steady solution is $\Psi=0$. Surprisingly, stable domains with well
defined plateaus can be obtained even in this regime, although the
time trace gets inverted every round-trip, thereby inducing a periodicity
close to twice the delay value $2\tau$, see Fig.~\ref{figtheo1}d-g).
Although the solution inversion is interesting in its own right, such
a behavior can be deduced intuitively from the singular mapping Eq.~(\ref{eq:Mapping})
setting $\kappa=1$. The striking result is that the domain walls
remain motionless, even for large $\beta$ values, see Fig.~\ref{figtheo1}d)
and Fig.~\ref{figtheo1}e) where the checkerboard pattern visually
disappears leaving only visible the transitions. We verified this
robustness with other kind of symmetry breaking terms e.g. $\gamma_{m}\Psi^{m}$
with $m=0$ and $m=4$. One can still construct approximate analytic
solutions of Eq.~(\ref{eq:PFPDE}) where an upward or a downward
kink must be complemented by an opposed kink \emph{at a spatial distance}
$x=1$ in order to fulfill the anti-periodic boundary condition Eq.~(\ref{eq:BC}).
A similar variational approach allows us to find the effective equation
of motion of an isolated wall, 
\begin{eqnarray}
\frac{dx_{i}}{dn} & \sim & \beta\int_{0}^{2}\left(\partial_{x}K_{+}+\partial_{x}K_{-}\right)\left(K_{+}+K_{-}\right)^{2}dx=0.\label{eq:P2domain}
\end{eqnarray}

In other words, by inverting at each round-trip, the walls experience
opposed drift velocities that cancel out explaining why similar results
can be found for other kind of symmetry breaking nonlinearities. It
was also demonstrated in \cite{N-PD-03} that the quadratic term cancels
out of Eq.~(\ref{eq:PFPDE}) even if it is large, see Eqs.~(A.17-18)
in the appendix~A of \cite{N-PD-03} for details. Higher order nonlinearities
in Eq.~(\ref{eq:PFDDE}) potentially as large as $\gamma_{m}\sim\varepsilon^{3-m}$
may enters Eq.~(\ref{eq:PFPDE}), but would cancel out similarly
in Eq.~(\ref{eq:P2domain}). We conclude that P2 delayed systems
are robust candidates for information storing as they are impervious
to most experimental imperfections leading to a coarsening.

\begin{figure}
\begin{centering}
\includegraphics[bb=10bp 0bp 611bp 328bp,clip,width=1\columnwidth]{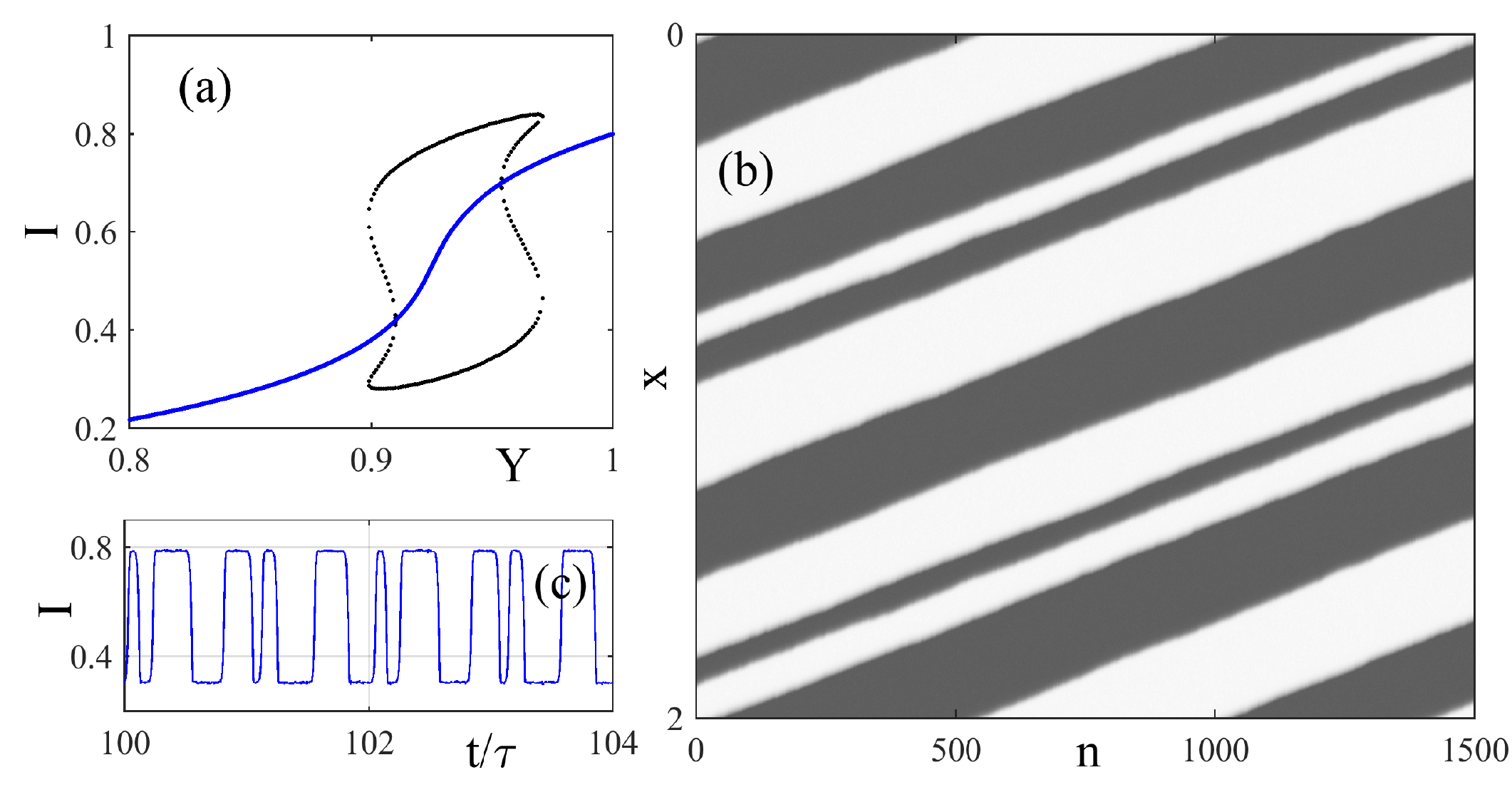}
\par\end{centering}

\caption{{\footnotesize{}(color online): a) Non linear output intensity $I=|E|^{2}$
as a function of the injected field $Y$ in blue. The black dots represents
the amplitude of the upper and lower plateau in the P2 regime. The
space-time diagram of the intensity in panel b) is folded with a period
$T=2\tau+0.1$ and shows the stability of the domain walls, even in
the presence of noise, see also panel c).} {\footnotesize{}Parameters
are $\alpha=2$, $J=-0.1$, $\Delta=2.3$, $\tilde{\eta}=0.1$ and
$\Omega=\pi$.}}
\label{figtheo2}
\end{figure}

We study the case of an injected semiconductor laser below threshold
subject to optical feedback, as domain stabilization in telecommunication
lasers is very interesting also from an applicative point of view.
Close, yet below, threshold and in the limit of weak optical feedback,
weak optical injection, small detuning and large delay, one can reduce
the standard single-mode rate equations of the semiconductor laser
to the following delayed Ginzburg-Landau equation 
\begin{eqnarray}
\hspace{-0.7cm}\dot{E} & = & \left(1+i\alpha\right)\left(J-\left|E\right|^{2}\right)E+i\Delta E+Y+\tilde{\eta}e^{-i\Omega}E\left(t-\tau\right)\label{eq:CA}
\end{eqnarray}
see for instance \cite{GJT-NC-15} for details. Here, $J$ denotes
the deviation from threshold of the bias current. The detuning between
the frequency of the injection field $\omega_{Y}$ of amplitude $Y$
(chosen real by definiteness) and the free running frequency of the
laser scaled by the photon lifetime is $\Delta$, whilst $\tilde{\eta}$
and $\Omega=\omega_{Y}\tau$ are the amplitude and phase of the delayed
optical feedback, respectively. At steady state ($\dot{E}=0$), the
output power $\left|E\right|^{2}$ as a function of $Y$ can present
a bistable S-shape response for some parameters. We work close to
the onset of bistability where the transition from low to high power
has a sigmoid shape, see the blue line in Fig.~\ref{figtheo2}a).

Assuming that a P2 solution evolves between two plateaus whose values
are $E_{\pm}$, it yields a system of equations whose solution is
represented by the black dots in Fig.~\ref{figtheo2}a,b). Such a
diagram suggests that these P2 regimes are nascent from two saddle-node
bifurcations of limit cycles and are also connected to the stationary
solution (blue line) by two Andronov-Hopf bifurcations. However these
bifurcations are subcritical meaning that the equivalent PDE would
certainly take the form of a subcritical cubic-quintic Ginzburg-Landau
equation. Notice that for dimensional systems with a higher number
of variables, as in Eq.~(\ref{eq:CA}), additional instabilities
like the Eckhaus mechanism \cite{WY-PRL-06} may arise. Yet, Eq.~(\ref{eq:CA})
can be reduced to a single scalar delayed equation for the phase in
the limit of weak feedback and injection using averaging methods \cite{N-PD-02}.
We show indeed in Fig.~\ref{figtheo2}b,c) that this system is capable
storing patterns of domain walls. 

\begin{figure}
\begin{centering}
\includegraphics[width=1\columnwidth]{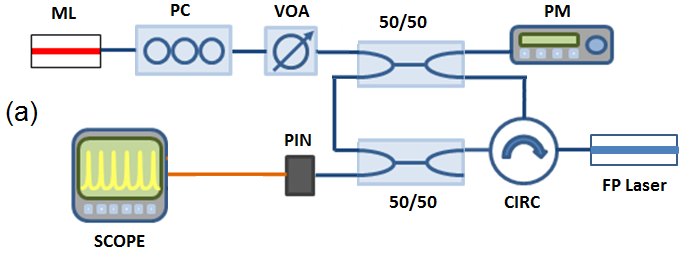}
\par\end{centering}

\begin{centering}
\includegraphics[width=0.5\columnwidth]{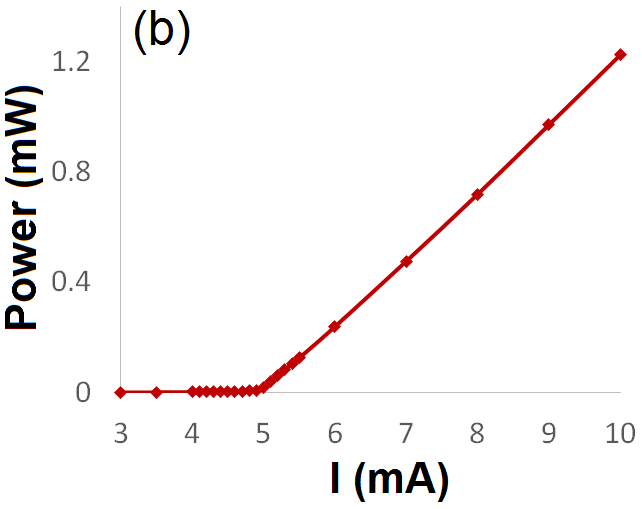}\includegraphics[width=0.5\columnwidth]{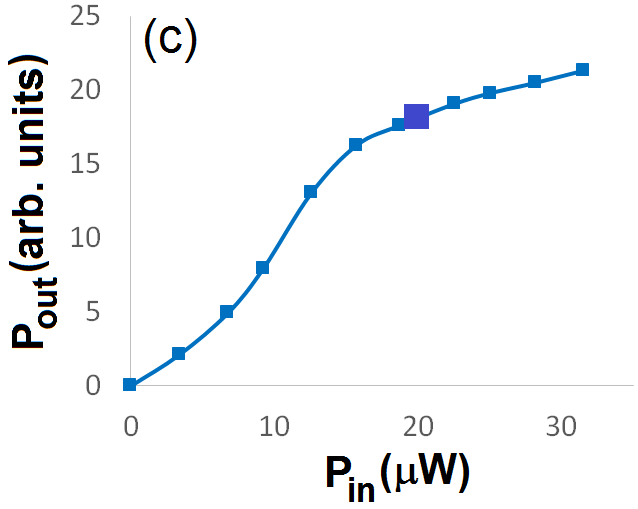}
\par\end{centering}

\caption{{\footnotesize{}(color online) (a) Experimental Setup. Legend: ML:
Master Laser; PC: Polarization Controller; VOA: Variable Optical Attenuator;
PM: Power Meter; FP: Fabry-Perot; CIRC: Circulator; PIN: $12\,$GHz
Photodetector; SCOPE: $13\,$GHz Oscilloscope. (b) LI-curve of the
SL at $T=293\,$K. (c) I/O power relationship of the SL under external
injection with $\Delta f=-10\,$GHz and $I=4.85\,$mA.}}

\label{setup}
\end{figure}

The experimental setup is represented in Fig.~\ref{setup}. A 1310~nm
Fabry-Perot Laser (Slave Laser, SL) as used in telecommunication systems
is subject simultaneously to external optical injection and delayed
feedback after a round-trip time $\tau=65.4\,$ns. The optically injected
signal was generated with a tunable laser (Master Laser, ML). A polarization
controller and a variable optical attenuator were respectively included
after the ML to control its polarization state and optical power level.
Figure~\ref{setup}b) plots the light-current (LI) curve of the solitary
SL, showing a threshold current of $I_{th}=4.92\,$mA. Figure~\ref{setup}c)
depicts the SL's input/output power relationship when subject solely
to external injection. The device was biased below threshold with
a current of $I=4.85\,$mA (i.e. $I=0.985\, I_{th}$) and an initial
frequency detuning ($\Delta f=f_{inj}-f_{FP}$) equal to $-10\,$GHz
was set between the frequencies of the injected signal and the resonance
frequency of one of the SL modes. Figure~\ref{setup}c) illustrates
the achievement of a gradual nonlinear switching response as the injection
strength is increased from 0 to $31.5\,\mu$W in agreement with the
results of Fig.~\ref{figtheo2}a).

\begin{figure}[t]
\begin{centering}
\includegraphics[bb=0bp 0bp 440bp 320bp,clip,width=1\columnwidth]{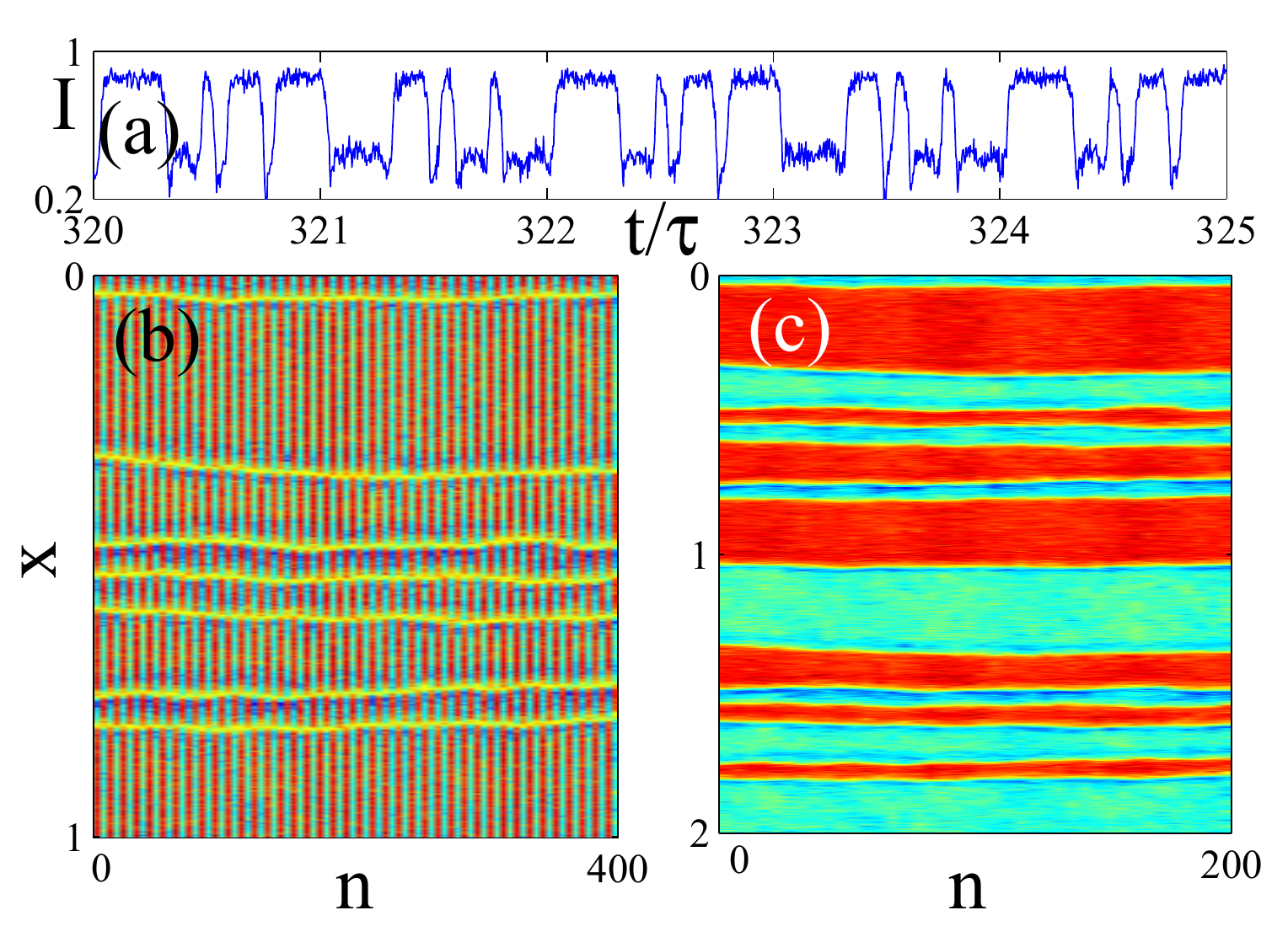}
\par\end{centering}

\caption{{\footnotesize{}(color online). Temporal trace of the laser intensity
a) and space-time diagram with folding parameter $\tau$ b) and $2\tau$
c).}}
\label{figexp1}
\end{figure}

The SL was subsequently subjected to simultaneous optical injection
from the ML (input power $P_{inj}=20\,\mu$W) and delayed feedback.
In this situation the system was operated in the sigmoid range as
indicated by the blue square in Fig.~\ref{setup}c). The phase of
the feedback can be changed by a small detuning of the injection field
frequency. The large values of the time delay allows considering these
two parameters independent.

Figure~\ref{figexp1}a) plots time trace over 5 round-trips ($327\,$ns).
Figure~\ref{figexp1}a) shows that a temporal pattern which is inverted
every round-trip is obtained at the device's output, in agreement
with the predictions of Fig.~\ref{figtheo1}g) and Fig.~\ref{figtheo2}c).
The equivalent space-time diagrams for the time series of Fig.~\ref{figexp1}a)
when the folding parameter is set approximately to $\tau$ and $2\tau$
are plotted respectively in Fig.~\ref{figexp1}b) and Fig.~\ref{figexp1}c)
over a time window of $400\,$round-trips (i.e. $\sim26\,\mu$s).
The figures~\ref{figexp1}b-c) demonstrate the experimental achievement
of anti-periodic dynamical regimes since peaks-troughs alternate every
round-trip, see Fig.~\ref{figexp1}b). More importantly, the results
also demonstrate the formation of stable domains of arbitrary size.
The existence of various noise sources in the system induces a slow
drift of the walls as seen in Fig.~\ref{figexp1}c), but not a coarsening
if the walls remain at a sufficient large distance, in good agreement
with theory.

\begin{figure}[t]
\begin{centering}
\includegraphics[bb=0bp 0bp 555bp 281bp,clip,width=1\columnwidth]{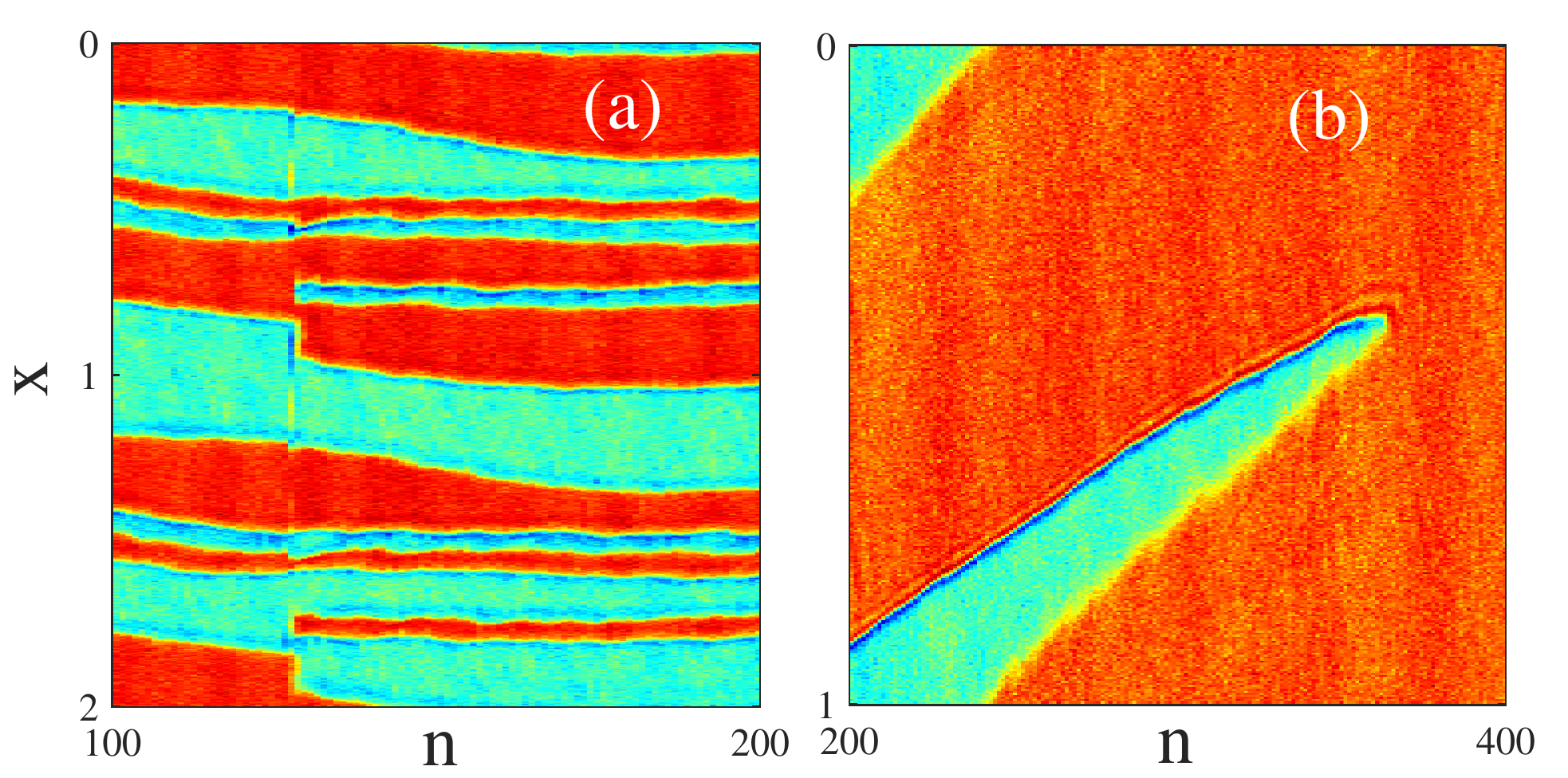}
\par\end{centering}

\caption{{\footnotesize{}(color online). Same as Fig.~\ref{figexp1}. A new
pair of domain wall is nucleated at $n=128$ in panel a) and remain
stable while for different parameters bistable phase coarsening occurs.}}
\label{figexp2}
\end{figure}

Noteworthy, we observed the nucleation of domain wall pairs after
the arrival of an electrical perturbation into the system as shown
in Fig.~\ref{figexp2}a) at $n=128$. Additionally, a small change
of the injection parameters allows to pass into the weakly bistable
regime and to observe in Fig.~\ref{figexp2}b) the coarsening mechanism
of the P1 solutions, as in \cite{GMZ-EPL-12}.

In conclusion, we presented in this manuscript how a general property
of delayed systems can be harnessed to prevent the domain coarsening
in symmetry broken delayed systems thereby allowing the storing of
information. Such anti-periodicity has no equivalent in real spatially
extended systems. We evidence the existence of stable domains in the
coherent output of a semiconductor laser with optical feedback. These
results offer exciting prospects for the controllable encoding of
information.
\begin{acknowledgments}
J.J. acknowledges useful discussion with S. Balle as well as financial
support from Ramón y Cajal program, and project RANGER (TEC2012-38864-C03-01).
A.H. thanks Prof. A. Kemp for lending the oscilloscope used in the
experiments and financial support from the Stratchlyde Chancellor's
Fellowships Programme: Starter Grant (Ref 12431DP2425AOOO), Institute
of Photonics, University of Strathclyde

\end{acknowledgments}

\end{document}